# Ni/Ni$_3$C Core-Shell Nanochains and Its Magnetic Properties: One-Step Synthesis at low temperature

Wei Zhou,[†] Lin He,[‡] Rongming Wang,[†] Kun Zheng,[§] Lin Guo,*[†] Chinping Chen*[‡], Xiaodong Han*[§] and Ze Zhang[§]

[†]School of Materials Science and Engineering, School of Science, Beijing University of Aeronautics and Astronautics, Beijing 100083, [‡] Department of Physics, Peking University, Beijing 100871 and [§]Institute of Microstructure and Properties of Advanced Materials, Beijing University of Technology,Beijing 100022, P. R. China

* To whom correspondence should be addressed. E-mail: guolin@buaa.edu.cn; cpchen@pku.edu.cn; xdhan@bjut.edu.cn.

[†]School of Materials Science and Engineering, Beijing University of Aeronautics and Astronautics. [‡]Department of Physics, Peking University. [§]Institute of Microstructure and Properties of Advanced Materials, Beijing University of Technology.

**ABSTRACT**  One-dimensional Ni/Ni$_3$C core-shell nanoball chains with an average diameter by around 30 nm were synthesized by means of a mild chemical solution method using a soft template of trioctylphosphineoxide (TOPO). It was revealed that the uniform Ni nanochains were capped with Ni$_3$C thin shells by about 1~4 nm in thickness and each Ni core consists of polygrains. The coercivity of the core-shell nanochains is much enhanced (600 Oe at 5 K) and comparable with single Ni nanowires due to the one-dimensional shape anisotropy. Deriving from the distinctive structure of Ni core and Ni$_3$C shell, this architecture may possess a possible bi-functionality. This unique architecture is also useful for the study on the magnetization reversal mechanism of one-dimensional magnetic nanostructure.

**Introduction.** Nanoscale magnetic materials such as Fe, Co, Ni have attracted much attention because of their unique magnetic, catalytic and optical properties with promising applications in magnetic sensors, high-density magnetic records, catalysts, etc.[1–4] In addition, some of the basic issues about magnetic and optical phenomena in one-dimensional systems have also been addressed with these materials.[4,5] Accordingly, a series of one-dimension (1D) nickel nanostructures including nanotubes, nanowires, nanorods, nanochains and nanoarrays have been synthesized.[6–13]

A variety of one-dimensional (1D) core-shell structures were prepared to enhance the multi-functionality of these materials, and they have exhibited good characters in improving luminescent efficiency, magnetic property and as field effect transistors.[14–17] Many methods of preparation in high temperature, including a laser ablation method, a high-temperature route and a carbothermal reduction method, have been used to synthesize the core-shell nanomaterials.[18–20] In addition, several solution-based methods have emerged to generate nanocables having a core/shell structure with a metal core and a sheath of different materials, including polymer, metal, semiconductor and insulator, at a relatively low temperature.[21–24] However, it remains a challenge to prepare the 1D Ni/Ni$_3$C core-shell structure, which has the metallic, soft ferromagnetic Ni in the core and the nonmagnetic Ni$_3$C as the outer protective sheath. Herein, we report a simple one-step chemical solution method to synthesize the core-shell structured Ni/Ni$_3$C nanochains at a low temperature. The structure, substructures and the morphologies as well as the correlated magnetic properties of the nanochains were studied systematically. The mechanisms and a simplified model of formation for the Ni/Ni$_3$C nanochains were proposed.

**Experimental Section.** In a typical experiment, all the reagents were analytical grade and used without further purification. A solution was made by dissolving 0.119 g of $NiCl_2 \cdot 6H_2O$ and 0.309 g of trioctylphosphineoxide (TOPO) in the solvent of 60 ml ethylene glycol (EG) at room temperature. After a strong stirring for half an hour, a light-green color with a little turbid appeared in the solution. 1 ml of 50wt % hydrazine monohydrate was dropwise added into the solution until the color turned into light purple and then light pink. The stable pink solution was heated to the boiling point of glycol (197 °C) and refluxed for 5 hours under a vigorous magnetic stirring. Finally, the resulting black precipitates were separated and washed with distilled water and ethanol for several times. The as-synthesized products were then dried at 60 °C.

The crystal structure of the as-prepared product were characterized by X-ray powder diffraction (XRD) using a Rigaku Dmax 2200 X-ray diffractometer with Cu Kα radiation (λ=0.1542 nm). The XRD specimens were prepared by means of flattening the powder on the small slides. The general morphologies of the synthesized nanomaterials were studied by a field-emission gun (FEG) scanning electron microscope (Hitachi S-4300, 5 kV) with the samples obtained from the thick suspension dropping on the glass platelets. Transmission electron microscopy (TEM) investigations were carried out with a JEOL 2010 FEG ultrahigh resolution scanning transmission electron microscope (STEM). The point resolution of the microscope is about 0.19 nm. Under STEM mode, the probe size is less than 0.2 nm. The energy resolution of EELS analysis is about 1.2 eV. The operating voltage for all of the TEM and STEM experiments were 200 KV. The as-grown nanochains were dispersed in ethanol and dropped onto a holey carbon film supported on a copper grid with drying in air. Magnetic properties of the nanochains were measured using a Quantum Design SQUID-based magnetometer (MPMS system).

**Results and Discussion.** The Ni nanoball-chains covered with a shell layer of $Ni_3C$ were synthesized. Figure 1a reveals the general morphology of the synthesized nanomaterials. These nanomaterials are with ample spherical particles forming in short chainlike structures. The TEM bright field image as shown in Figure 1b demonstrates a view for the as-produced nanoball chains. It reveals that the particles are structurally connected together, forming nanochain networks. The nanoballs are found to have a uniform size with the diameter of 30±10 nm. Figure 1c shows a high magnification TEM bright field image for a single nanoball which reveals clearly a core-shell structure. The subsequent sections will demonstrate that this core-shell nanoball consists of a Ni core and a $Ni_3C$ shell of 1~4 nm in thickness. Figure 1d is a selected area electron diffraction (SAED) pattern taken from Figure 1b. The diffraction rings can be indexed with the face centered cubic (FCC) Ni and the rhombohedra (RH) $Ni_3C$ as shown in Figure 1d. The X-ray diffraction (XRD) pattern shown in Figure 1e confirms the above results. The main diffraction peaks marked with diamonds correspond to reflections of face-centered cubic (FCC) Ni with lattice parameter a = 0.3514 nm (JCPDS 04-0850). The other peaks marked with triangles can be assigned to be (110), (006), (113), (116), (300) and (119) planes of $Ni_3C$ (JCPDS 77-0194). The peak of $(111)_{Ni}$ overlays with the peak of $(113)_{Ni3C}$. Both of the SAED pattern and the XRD result show well crystalline characters of Ni and $Ni_3C$ phases.

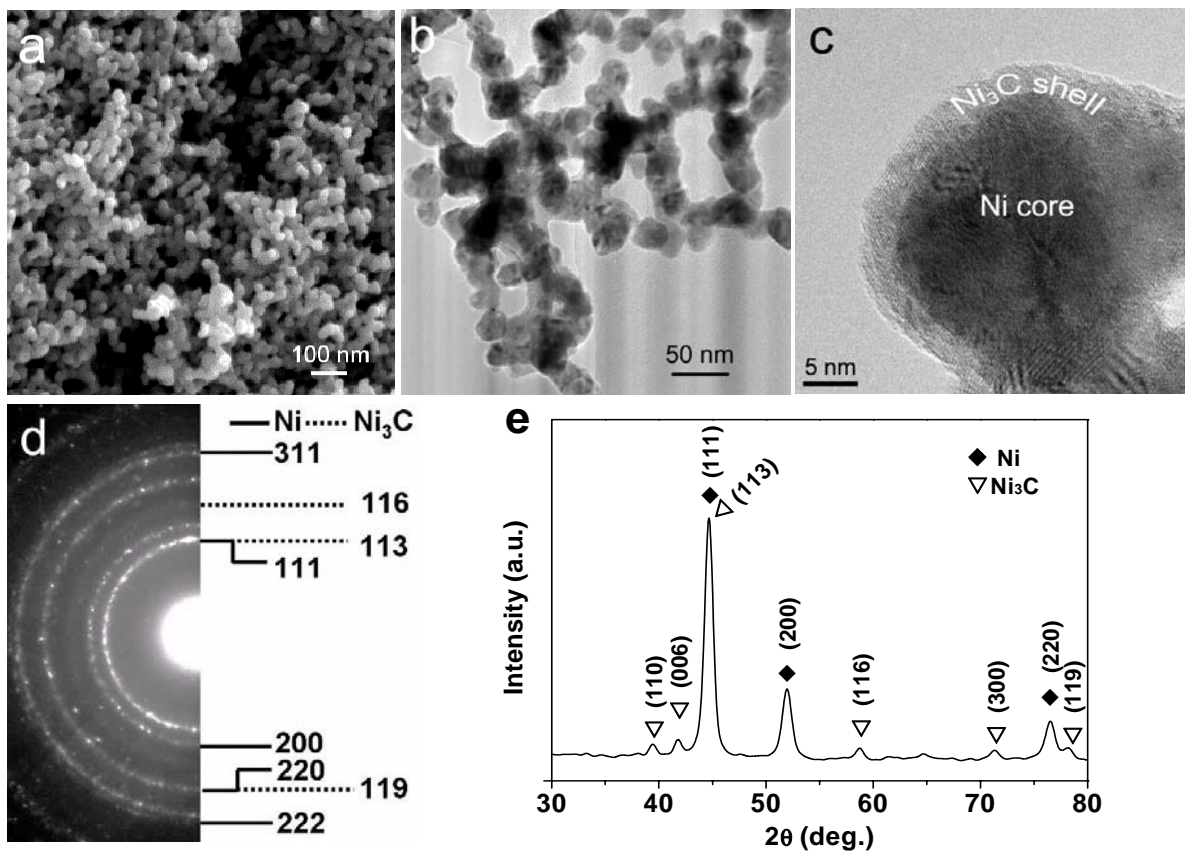

**Figure 1.** (a) SEM overview image of $Ni/Ni_3C$ nanochains. (b) Bright field TEM image of the chainlike network. (c) High magnification TEM image of a single nanoball from the top of a chain, showing the core-shell structure. (d) SAED pattern from several nanoballs of Figure 2b. (e) XRD pattern of the as-grown product.

The structure and substructures of the Ni/Ni₃C nanoball chains were further studied at atomic scale. Figure 2a shows a low magnification image of several connected nanoballs. Figure 2b is a HRTEM image taken from the cyan framed region of Figure 2a. The elongated nanoball consists of a core about 10 nm in diameter and a shell about 4nm in thickness. A Fast Fourier Transformation (FFT) pattern, shown as Figure 2c, is arranged at the right bottom in Figure 2b. It shows a well crystalline feature of the core-shell structure with the three sets of diffraction patterns. The diffraction spots marked as 1, 2 and 3 in Figure 2c for the FFT diffraction pattern correspond to the lattice zones 1, 2 and 3 in Figure 2b, respectively. The diffraction spots 1 and 2 can be assigned to $(110)_{Ni_3C}$, while 3 to $(111)_{Ni}$.

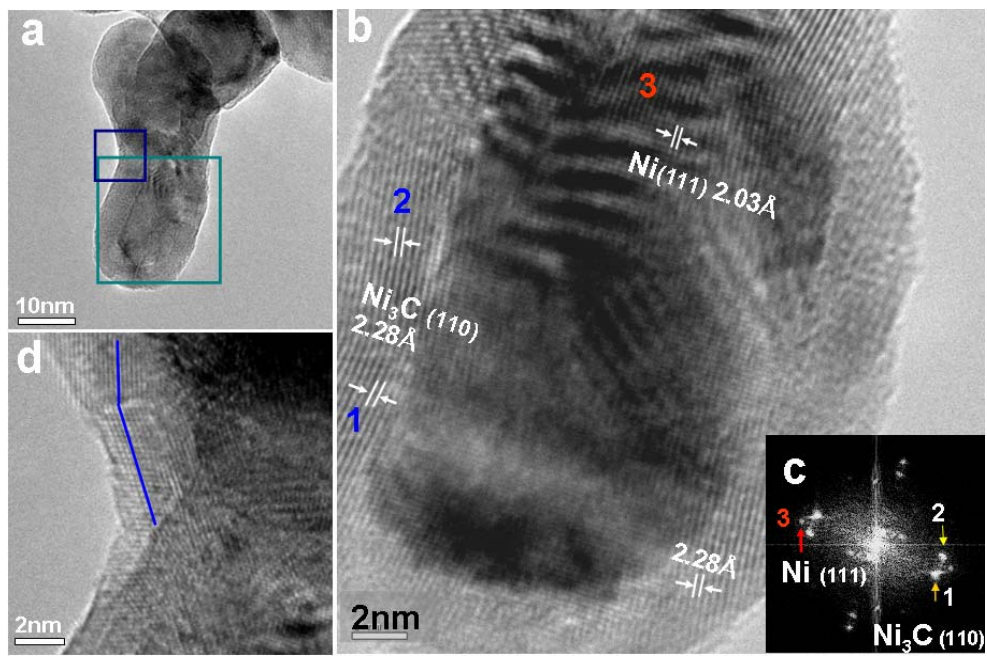

**Figure 2.** (a) HRTEM image of a particle at the tip of a chain. (b) Magnified image of the cyan square in Figure 2a. (c) FFT pattern corresponds to Figure 2b. The diffraction spots 1, 2 and 3 correspond to the lattices zone 1, 2 and 3 in Figure 2b. (d) Enlarged part of the blue square in Figure 2a.

As revealed in Figure 2b, the shell's lattice zone "1" alters its orientation to be along lattice zone "2" to match the inner Ni lattice structure, therefore, to minimize the lattice mismatch energy. Figure 2d is a magnified HRTEM image of the blue-framed region in Figure 2a taken near the connecting region of two contiguous Ni nanoparticles. Obviously, both of the core and the shell's lattice fringes consecutively cross the two nanoparticles without interruption by an apparent boundary. The blue lines draw the lattices of $Ni_3C$ shell which show the varying of the crystalline orientation. This indicates that the $Ni_3C$ shell grew on the surface of Ni core by an epitaxial way in which, due to the ball-shaped Ni core, the $Ni_3C$ shell appeared to be with a polycrystalline structure to minimize the localized lattice-mismatch energy (Please note that even in the single crystalline region of Ni core, the $Ni_3C$ still appears as polycrystalline feature). This suggests a post-growth process of the $Ni_3C$ shell over the Ni polycrystalline core.

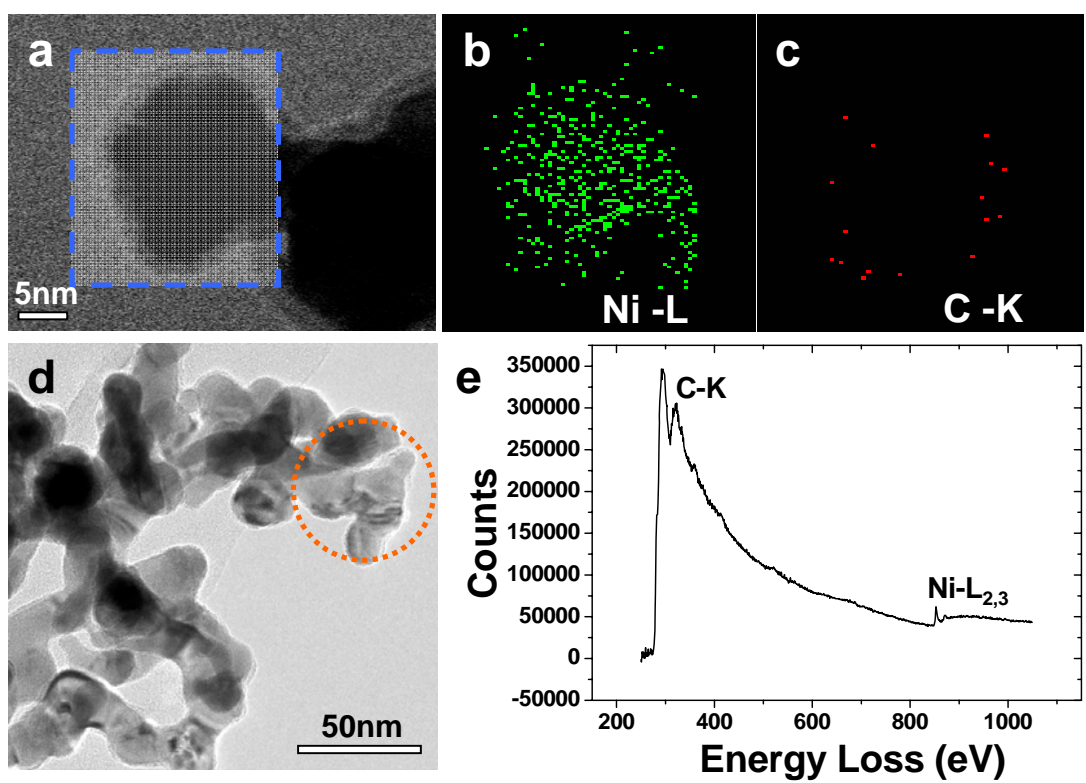

**Figure 3.** (a) TEM image and the corresponding EDS elemental maps of a nanoball, revealing the spatial distribution of (b) nickel and (c) carbon elements. (d) TEM image for the scanning region under the EELS analysis. (e) EELS spectrum showing C-K edge and Ni-L edge.

The shell component of the nanoball is demonstrated to be $Ni_3C$ by energy dispersion spectroscopy (EDS) and electron energy loss spectroscopy (EELS). Figure 3a is an image for a nanoparticle with the Ni/Ni$_3$C core-shell structure. Figures 3b and 3c show EDS elemental mapping scan results of the nanoball taken from the framed region of Figure 3a for elements Ni and C, respectively. Figure 3c shows a clear carbon shell structure giving the distribution of C element. To reveal the carbon bonding structure in the shell, EELS analysis was conducted. Figure 3d shows the region within which the spectrum was taken, and Figure 3e shows the corresponding EELS spectrum. The carbon K-edge and nickel L-edge are revealed in the spectrum at 291 eV and 855 eV, respectively. The carbon K-edge shows strong and sharp δ bond character [25,26] which is attributable to the Ni-C bonds. It has the similar feature with the diamond structure without any trace of π bonds. [25,26] The X-ray results and the polycrystalline diffraction rings in Figure 1 excludes the possibility of the out-shell structure being diamond-like carbon. The half-width of the K-edge peak is sharp and narrow. It indicates an ordered structure of carbon and a constant distance of the C-Ni bonding.

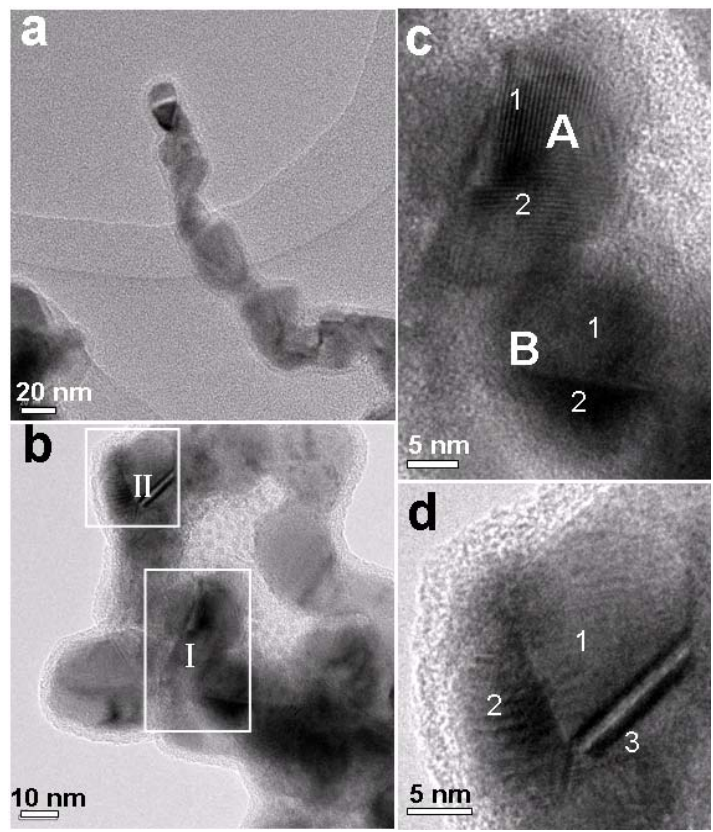

**Figure 4.** (a) A typical core-shell Ni nanochains. (b) HRTEM image showing the core-shell structure of the nanochain and the polycrystalline feature of the Ni cores. (c) An enlarged image for region I in (b). (d) An enlarged image for region II in (b).

Following the above microstructural analysis, we finally clarify most Ni core is with a polycrystalline feature. Figure 4a shows a typical core-shell nanochain. Figure 4b reveals the polycrystalline features of the Ni core. Figures 4c and 4d are the images with higher resolution for the framed areas I and II in Figure 4b, respectively. Figure 4c shows two Ni nanoballs (A and B) connects with each other inside a common $Ni_3C$ shell. Both of them consist of two nanograins, marked as 1 and 2. On the other hand, the nanoball in Figure 4d consists of three nanograins, marked as 1, 2, and 3, inside a common $Ni_3C$ shell. It can be concluded that each shell of the nanoball contains 1~3 Ni monocrystal grains. According to the characterizations by the EDS, EELS, and XRD analyses along with the TEM and HREM observations, it is concluded that the nanochains are with the $Ni/Ni_3C$ core-shell structures. The inner Ni cores, each of which consists of 1~3 nanograins, are in connection with each other structurally with a common $Ni_3C$ shell layer grown in the outer surface.

The temperature dependent magnetization, $M(T)$, curves shown in Figure 5(a) were measured by the zero-field-cooling and field-cooling (ZFC and FC) modes from 5 K to 380 K. To perform the ZFC measurement, the procedure was to cool the sample under zero applied field down to 5 K, and then applied a field of 90 Oe for data collection in the warming process. For the FC curve, the procedure was the same as in the ZFC measurement, except that the sample was cooled with the presence of an applied magnetic field of 20 kOe. These two curves separate widely from each other, indicating the presence of a magnetic anisotropy barrier. The blocking temperature corresponding to this anisotropy is much higher than 380 K. Interestingly, for a comparison with the result from the pure Ni nanochain with a free surface,[13,27] the freezing peak at $T \sim 13$ K attributed to the surface spin glass state is completely suppressed with the present $Ni/Ni_3C$ sample.[27] This indicates that the thin $Ni_3C$ shell layer serves not only as a chemically inner encapsulation layer but also as a magnetically surface modification layer.

Figure 5b shows the *M*(*H*) data measured at 5 K, 100 K and 300 K. The saturation magnetization determined in the high field region at 300 K is 35.4 emu/g, ~0.37 $\mu_B$ per Ni atom. It accounts for 61% of the corresponding bulk value, ~ 0.606 $\mu_B$/Ni. The reduction of the saturation magnetization is most likely arising from the existence of the nonferromagnetic $Ni_3C$ [28] shell layer. By assuming a spherical core-shell structure with an average outer diameter of 30 nm, the reduction in the observed saturation magnetization is attributable to an averaged $Ni_3C$ shell thickness of about 2.3 nm. This is reasonably consistent with the result of the shell thickness, 1~4 nm, observed by the HRTEM investigation.

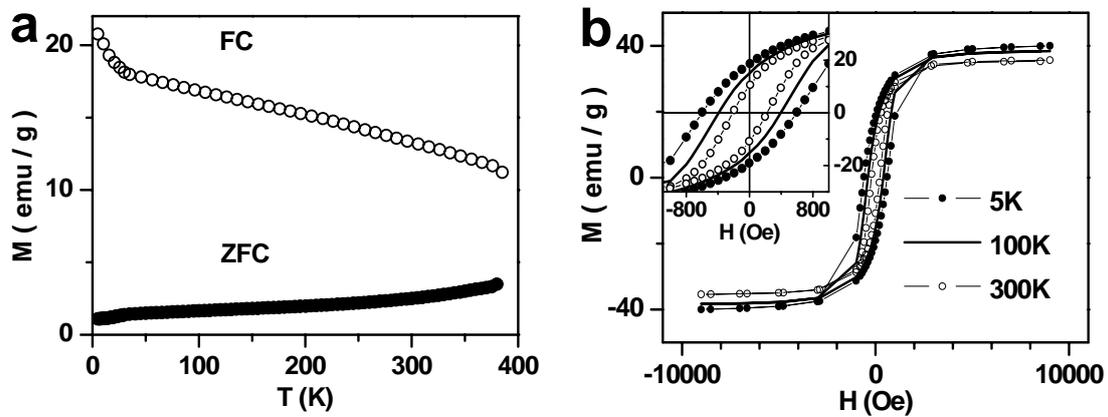

**Figure 5.** (a) ZFC and FC magnetization, *M*(*T*), measured with the applied field of 90 Oe. (b) *M*(*H*) data measured at 5 K, 100K and 300K. The inset shows the open hysteresis in the low field region.

With decreasing temperature, the saturation magnetization increases slightly, as shown in Figure 5b. An open hysteresis loop was observed in the low field region. The coercivity field, $H_C$, at 5 K was determined as 600 Oe, which is similar to the result, ~ 600 Oe, obtained for the Ni nanowires with the diameters ranging from 40 to 100 nm and lengths up to 5 μm at *T* = 5 K.[29] The large coercivity is not unexpected due to the one-dimensional shape of the samples. The magnetization reversal mechanism in the present nanochain structure can be described by the fanning mode of magnetization reversal according to the "chain of spheres" model.[30] This is reasonable due to the fact that the diameter of the sphere, 25.4 nm (30 nm − 4.6 nm), is comparable to the coherence length of Ni, ~ 25 nm. With the fanning mode, the $H_C$ measured along the chain is expressed as,

$$H_{C,n} = \frac{\mu}{R^3}(6K_n - 4L_n) \qquad (1)$$

where $\mu$ and R are the dipole moment and the diameter of each sphere. The expression in the parenthesis takes into account for the dipolar interaction between the magnetic spheres in the chain based on the assumption of fanning mode. In particular, $K_n$ accounts for the dipolar interaction between each every pair of the magnetic particles, and $L_n$, between each odd-numbered and even-numbered pairs. The number, n, is for the sphere numbering in the chain.[30] This model considers the low temperature property without accounting for the thermal activation effect. The term $\mu/R^3$ can be expressed, using saturation magnetization per unit volume $M_S$, as $\mu/R^3 = (\pi/6) M_S$. Thus, by using the experimental value of $M_S$, the density of bulk Ni and assuming n =12, we obtain the values of $H_C$ as 485 Oe. By taking into account the correction factor for the randomly oriented effect in a powdered sample, the value of the coercivity is modeified as $H'_C \sim 1.1 H_C = 534$ Oe. This agrees with our experimental result, ~ 600 Oe, within 11%. The remanent magnetization at 5 K is about 19 emu/g. The corresponding remanent ratio is 0.475. This is in agreement with the prediction of fanning model, ~ 0.5, for the randomly oriented particles within 5%. With increasing temperature, both the coercivity and remanent magnetization decrease, as shown in the inset of Figure 5b, due to the thermal activation effect.[31] The coercivity and remanent magnetization are 390 Oe and 15 emu/g at 100 K, respectively. They become 217 Oe and 10.5 emu/g at 300 K.

Normally, $Ni_3C$ is obtained by physics methods with the conditions of high temperature,[32] and high pressure, such as mechanical alloying,[28,33] C-ion implantation into Ni,[34] and reaction between Ni and amorphous C films.[35] Recently, mild chemical solution methods have been applied to synthesize nanosized nickel carbide. For example, Leng et al. prepared pure $Ni_3C$ nanoparticles by thermal decomposition in solution at 529 K.[36] However, most Ni-based nanomaterials with the core-shell structure were reported with Ni/C [37,38,39] or $Ni_3C$/C [39,40] nanoparticles. To our best knowledge, the core-shell Ni/$Ni_3C$ nanochain structure has never been reported before. More significantly, in our present case, the Ni/$Ni_3C$ core-shell architecture was obtained by a one-step chemical solution method below 473 K.

The formation of the Ni/Ni$_3$C core-shell structure could be explained by the following simplified model and process. First, as shown in Figures 6a, Ni particles form in several minutes in the solution with the presence of TOPO (C$_{24}$H$_{51}$OP) at the boiling point of glycol. It has been known that TOPO is a surface modifying agent for preparing Ni nanomaterials.[9,41] Under the surface modification effect of TOPO and the magnetic dipolar interaction of the Ni grains, the one-dimensional nanochain structure was evolved rapidly as illustrated in Figure 6b.

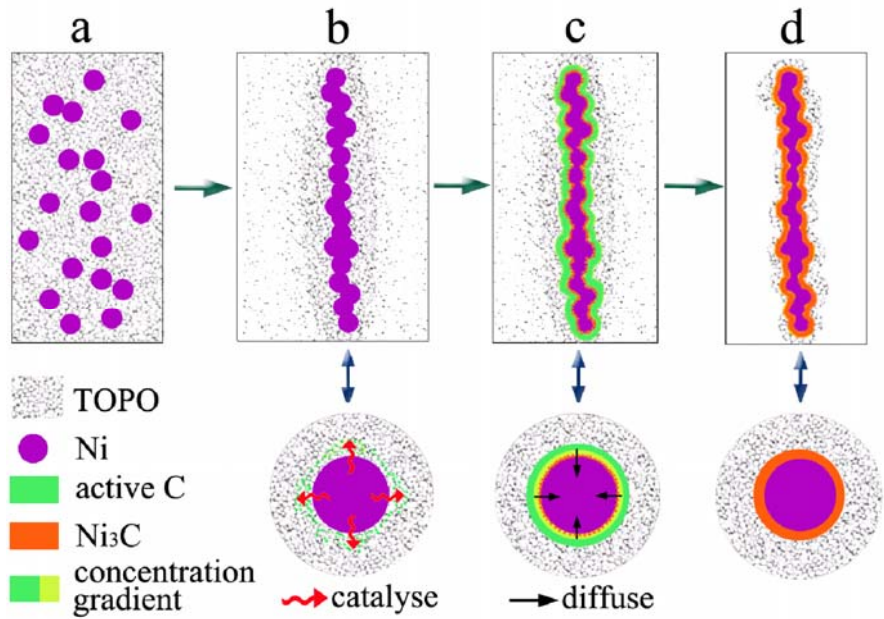

**Figure 6.** Illustration of the formation of the Ni/Ni$_3$C core-shell nanochain. (a) the formation of Ni particles in TOPO solution; (b) the formation of nano Ni-chain covered with TOPO is shown at the top. An enlarged cross-sectional view is shown at the bottom. Catalyzed by Ni and come from TOPO, the free active carbon atoms started to form on the surface of nano Ni-chain, the red-snaked arrows indicate the catalyzation process; (c) the carbon atoms diffused into the Ni lattice and naturally introduced a concentration gradient shell of carbon-Ni. The top and bottom sets are the side view and the enlarged cross-sectional view, respectively. The black arrows indicate the diffusion process; (d) the formation of Ni3C on the Ni surface and core-shell structure of Ni3C-Ni. The top and bottom sets are the side-view and enlarged cross-sectional view, respectively.

Meanwhile, deriving from electrostatic interactions, TOPO molecules were adsorbed on the surface of the Ni chains. Similar to the formation of carbon by their low-temperature cleavage of alkanes,[42] the small Ni particles catalyze the formation of active carbon from the organic molecules (TOPO) on their surfaces (Figure 6b).[36,43] As the restricted forming rate of active C atoms at the temperature of 470K (the boiling point of EG), the formation of $Ni_3C$ (Figure 6d) is posterior to that of Ni chains (Figure 6b). Then, the active carbon atoms gradually diffuse from the surfaces into Ni nanochains driven by the concentration gradient of the carbon (Figure 6c),[36,44,45] forming the $Ni_3C$ interstitial compound. The prior-formed $Ni_3C$ shells were likely to prevent the formation of additional carbide by stopping the contact of the Ni atoms and the active carbon atoms.[44,45] This process made an intact Ni-core in the centre and formed this core-shell structure(Figure 6d), finally.

In summary, the $Ni/Ni_3C$ core-shell nanochains were formed with the surface modification of trioctylphosphineoxide (TOPO) at a low temperature (below 473K). This unique architecture may play bi-functionality deriving from the magnetic Ni core and the non-magnetic and oxidation resistant $Ni_3C$ shell. The $Ni_3C$ layer was formed by the active carbon atoms diffusing into the Ni particles on the surfaces. The mechanism could be used to guide the synthesis of similar core-shell nanomaterials. By magnetic measurements, the saturation magnetization of the nanochains is reduced sharply (accounting for 61% of the corresponding bulk value) due to the presence of the nonmagnetic $Ni_3C$. However, the coercivity is much enhanced (600 Oe at 5K) attributed to the one-dimensional morphology of the sample. Both of the coercivity and the remanent magnetization are in good agreement with the description of the fanning model at a low temperature. This unique architecture is useful to study the magnetization reversal mechanism for understanding the basic correlated magnetic physics.

**Acknowledgment.** Authors acknowledge the support from the National Natural Science Foundation of China (20673009, 50725208 & 50671003), SRFDP-20060006005, the program for New Century Excellent Talents in University (NCET-04-0164 & NCET-05009015299701), the National Basic

Research Program of China (2007CB935400 & 2006CB932300) and XDHAN thanks Beijing Education Committee Key Program (Beijing Nature Science Foundation Key Program)

**FIGURE CAPTIONS**

**Figure 1.** (a) SEM overview image of Ni/Ni$_3$C nanochains. (b) Bright field TEM image of the chainlike network. (c) High magnification TEM image of a single nanoball from the top of a chain, showing the core-shell structure. (d) SAED pattern from several nanoballs of Figure 2b. (e) XRD pattern of the as-grown product.

**Figure 2.** (a) HRTEM image of a particle at the tip of a chain. (b) Magnified image of the cyan square in Figure 2a. (c) FFT pattern corresponds to Figure 2b. The diffraction spots 1, 2 and 3 correspond to the lattices zone 1, 2 and 3 in Figure 2b. (d) Enlarged part of the blue square in Figure 2a.

**Figure 3.** (a) TEM image and the corresponding EDS elemental maps of a nanoball, revealing the spatial distribution of (b) nickel and (c) carbon elements. (d) TEM image for the scanning region under the EELS analysis. (e) EELS spectrum showing C-K edge and Ni-L edge.

**Figure 4.** (a) A typical core-shell Ni nanochains. (b) HRTEM image showing the core-shell structure of the nanochain and the polycrystalline feature of the Ni cores. (c) An enlarged image for region I in (b). (d) An enlarged image for region II in (b).

**Figure 5.** (a) ZFC and FC magnetization, $M(T)$, measured with the applied field of 90 Oe. (b) $M(H)$ data measured at 5 K, 100K and 300K. The inset shows the open hysteresis in the low field region.

**Figure 6.** Illustration of the formation of the Ni/Ni$_3$C core-shell nanochain. (a) the formation of Ni particles in TOPO solution; (b) the formation of nano Ni-chain covered with TOPO is shown at the top. An enlarged cross-sectional view is shown at the bottom. Catalyzed by Ni and come from TOPO, the free active carbon atoms started to form on the surface of nano Ni-chain, the red-snaked arrows indicate the catalyzation process; (c) the carbon atoms diffused into the Ni lattice and naturally introduced a concentration gradient shell of carbon-Ni. The top and bottom sets are the side view and the enlarged

cross-sectional view, respectively. The black arrows indicate the diffusion process; (d) the formation of Ni3C on the Ni surface and core-shell structure of Ni3C-Ni. The top and bottom sets are the side-view and enlarged cross-sectional view, respectively.

**SYNOPSIS TOC**

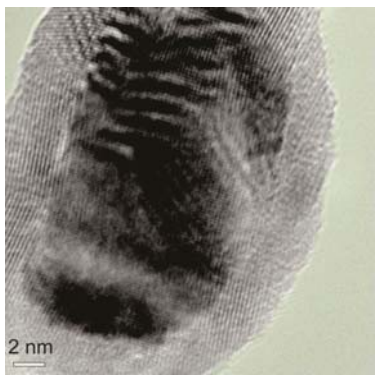

One-dimensional Ni/Ni$_3$C core-shell nanochains were synthesized by a one-step chemical solution method using a soft template of trioctylphosphineoxide (TOPO) at a low temperature (below 473K). This special structure is ideal to protect Ni nanochains from oxidation and to study the magnetization reversal mechanism.